\documentclass[aps,twocolumn,showpacs,preprintnumbers,nofootinbib,prd,superscriptaddress,10pt]{revtex4-1}

\makeatletter
\def\l@subsubsection#1#2{}
\def\l@subsubsubsection#1#2{}
\makeatother

\usepackage{comment}
\usepackage{graphicx,amssymb,amsmath,amsthm,amsfonts,epsfig,epsf,fixmath}
\usepackage[usenames]{color}
\usepackage{xcolor}
\usepackage{epstopdf}

\usepackage{aas_macros}
\usepackage{bm}
\usepackage{dcolumn}
\usepackage{lipsum}
\usepackage{latexsym}
\usepackage{rotating}
\usepackage{longtable}

\usepackage{enumerate}
\usepackage{tensor,multirow}
\usepackage{url}
\usepackage[linktocpage]{hyperref}

\defcitealias{Cannizzaro:2020uap}{Paper~I}

\begin{document}

\title{Plasma-photon interaction in curved spacetime.~II:\\collisions, thermal corrections, and superradiant instabilities}

\author{Enrico Cannizzaro}
\affiliation{Dipartimento di Fisica, ``Sapienza'' Universit\`a di Roma \& Sezione INFN Roma1, Piazzale Aldo Moro
5, 00185, Roma, Italy}
\author{Andrea Caputo}
\affiliation{School of Physics and Astronomy, Tel-Aviv University, Tel-Aviv 69978, Israel}
\affiliation{Department of Particle Physics and Astrophysics, Weizmann Institute of Science, Rehovot 7610001, Israel}
\author{Laura Sberna}
\affiliation{Max Planck Institute for Gravitational Physics (Albert Einstein Institute) Am Mu\"{u}hlenberg 1, 14476 Potsdam, Germany}
\author{Paolo Pani}
\affiliation{Dipartimento di Fisica, ``Sapienza'' Universit\`a di Roma \& Sezione INFN Roma1, Piazzale Aldo Moro
5, 00185, Roma, Italy}

\begin{abstract} 
Motivated by electromagnetic-field confinement due to plasma near accreting black holes, we continue our exploration of the linear dynamics of an electromagnetic field propagating in curved spacetime in the presence of plasma by including three effects that were neglected in our previous analysis: collisions in the plasma, thermal corrections, and the angular momentum of the background black-hole spacetime. We show that: (i) the plasma-driven long-lived modes survive in a collisional plasma except when the collision timescale is unrealistically small; (ii) thermal effects, which might be relevant for accretion disks around black holes, do not affect the axial long-lived modes; (iii) in the case of a spinning black hole the plasma-driven modes become superradiantly unstable at the linear level; 
 (iv) the polar sector in the small-frequency regime admits a reflection point due to the resonant properties of the plasma. Dissipative effects such as absorption, formation of plasma waves, and nonlinear dynamics play a crucial role in the vicinity of this resonant point.
\end{abstract}

\maketitle

\section{Introduction}

In Ref.~\cite{Cannizzaro:2020uap} (hereafter~\citetalias{Cannizzaro:2020uap}) we initiated an exploration of the linear dynamics of an electromagnetic~(EM) field propagating in a plasma within General Relativity. Previous studies simply assumed that a photon propagating in a plasma is dressed with an effective mass proportional to the plasma frequency, so that its dynamics in curved spacetime can be studied by solving the Proca equation for a massive spin-$1$ field.

In ~\citetalias{Cannizzaro:2020uap} we showed that the field equations are in general much more involved and richer than the effective Proca equation adopted in previous models (e.g.,~\cite{Pani:2013hpa,Conlon:2017hhi,Dima:2020rzg}). In particular, focusing on a cold, collisionless, and nonrelativistic plasma around a nonspinning black hole~(BH), we showed that the propagating degrees of freedom are only two, as expected from plasma theory in flat spacetime. The system therefore admits only a subset of the quasibound states found in the Proca case~\cite{Rosa:2011my}, and in general the existing quasibound modes are quantitatively different.

The scope of this follow-up work is to relax some of the working assumptions of~\citetalias{Cannizzaro:2020uap}, in particular by considering: (i)~collision effects in the plasma, including a discussion of the resonant behavior in the polar sector; (ii)~thermal effects, which are relevant for the plasma in accretion disks around astrophysical BHs; (iii)~the angular momentum of the central BH, which might turn the quasibound states into unstable modes at the linear level due to superradiance~\cite{Pani:2012bp,Pani:2012vp,Brito:2015oca}.

The above effects depend on the typical electron density and temperature of the plasma surrounding a BH, so in the following it might be useful to estimate the typical scales involved in astrophysical environments. We shall consider BHs in a wide range of masses from stellar-origin to supermassive (in practice assuming simply that $M\gtrsim M_\odot$) and an arbitrary dimensionless spin parameter $\tilde a$ (such that $0\leq\tilde a<1$).  The typical temperature of a thin accretion disk at a distance $r$ from the BH is approximately~\cite{Abramowicz:2011xu}
\begin{equation}
T \simeq 5\times 10^{7}\alpha^{-1/4}\Big(\frac{M_{\odot}}{M}\Big)^{1/4}\Big(\frac{r}{M}\Big)^{-3/8} \text{K}\,, \label{Tdisk}
\end{equation}
where $\alpha\sim {\cal O}(1)$.
The typical electron density near the BH depends strongly on its mass and on the geometry of the accretion flow. For a $M\sim 10^6 M_\odot$ BH, the particle density can be as high as $n_e\approx 10^{21}\,{\rm cm}^{-3}$ ($n_e\approx 10^{14}\,{\rm cm}^{-3}$) for geometrically thin (thick) accretion disks at the Eddington mass accretion rate~\cite{Abramowicz:2011xu,Barausse:2014tra}. For much lower accretion rates, or for quasispherical accretion, the electron density is much smaller, 
$n_e\approx 10^{4} \,{\rm cm}^{-3}$ or lower, depending on the ambient density at infinity (see, e.g.,~\cite{Caputo:2021efm}).

In the following we shall be mostly interested in configurations in which the plasma frequency,
\begin{equation}
   \omega_{\rm pl}= (n_e e^2/m_e)^{1/2},\label{plasmafreq}
\end{equation}
is smaller than the inverse size of the BH (we use $G=c=1$ units throughout), i.e. 
\begin{equation}
    \omega_{\rm pl} M\sim 0.25 \left(\frac{M}{M_\odot}\right)\left(\frac{n_e}{1\,{\rm cm}^{-3}}\right)^{1/2}\lesssim1\,,
\end{equation}
which limits the magnitude of the electron density for a given BH mass.

In this work, we show that collision and thermal effects do not affect significantly the quasibound states in the astrophysical environments of interest,
and that the spectrum can becomes superradiantly unstable at a linear level when the BH rotates. In general, the spectrum deviates from the one predicted in Proca theory ~\cite{Pani:2012vp, Pani:2012bp,Baryakhtar:2017ngi,Cardoso:2018tly,Dolan:2018dqv,Baumann:2019eav}, and a detailed analysis of perturbation theory in inhomogeneous plasmas is necessary to describe it.

\section{General equations}
\label{sec:setup}
An analysis of the dynamics of EM perturbations propagating in a (magnetised) plasma in curved spacetime was considered in Ref.~\cite{1981A&A....96..293B}, where the authors derived the perturbed system of equations in a plasma model consisting of a cold, nonrelativistic, collisionless fluid of electrons and ions.
Here we summarize this framework (see also \citetalias{Cannizzaro:2020uap}) and extend it to take into account plasma collisions and temperature effects.

\subsection{Cold and collisionless plasma}
\label{sub:general equation}

Let us denote the number density and four-velocity of the electrons (of mass $m_e$ and charge $e$) by $n_e$ and $u^{\mu}$, respectively, while $J^{\mu}$ stands for the ion current density. The system of differential equations for the plasma quantities reads
\begin{align}
\nabla_{\nu} F^{\mu\nu} &= e n_e u^{\mu} + J^{\mu}, \label{eq:Maxwell} \\
u^{\mu} \nabla_{\mu} u^{\nu} &= e/m_e F^{\nu}{}_{\mu}u^{\mu} ,\label{eq:momentum}\\
\nabla_{\mu}(nu^{\mu}) &= 0\,,\label{eq:last}\\
u^{\mu}u_{\mu} &= -1 \label{eq:normalisation} \,. 
\end{align} 
The above system simply consists of Maxwell's equations in the presence of sources, the electron momentum equation, the continuity equation for the electron fluid, and the normalisation of the electron's four-velocity. In particular, the momentum equation~\eqref{eq:momentum} relates the four-acceleration $a^{\nu}=u^{\mu} \nabla_{\mu} u^{\nu}$ to the external forces acting on the electrons, which for cold, collisionless plasmas only arise from the macroscopic EM field. As we shall discuss, we account for the gas temperature or electron-ion collisions by including extra terms in Eq.~\eqref{eq:momentum}.

We study the linearized dynamics of the system by introducing the small perturbations $\tilde{n}$, $ \tilde{u}^{\mu}$, $\tilde{F}_{\mu\nu}=(\partial_{\mu}\tilde{A}_\nu-\partial_{\nu}\tilde{A}_\mu)$, so that e.g. ${F}_{\mu\nu}={F}_{\mu\nu}^{\rm background}+\tilde{F}_{\mu\nu}$ and so on. We neglect higher-order perturbations of the plasma and EM field, as well as any perturbation of the background metric $g_{\mu\nu}$ (since the gravitational backreaction of these fields is small). We also neglect perturbations of the ions, since they will be suppressed with respect to those of the electrons by a factor $\propto m_e/ m_{\rm ion} \ll 1$.

We introduce the effective metric tensor 
\begin{equation}
h_{\mu\nu} = g_{\mu\nu} + u_{\mu}u_{\nu},
\end{equation}
which projects vectors and tensors onto hypersurfaces whose normal vector is the electron four-velocity. 
A generic tensor can be decomposed in the direction of the four-velocity and along the orthogonal directions by contracting it with the four-velocity and with the effective metric, respectively.
Performing the decomposition of the four-velocity gradient leads to 
\begin{equation}
\nabla_{\mu} u_{\nu} =\nabla {}_{(\mu}u_{\nu)} + \nabla {}_{[\mu}u_{\nu]}  ={\omega_{\nu}}_{\mu} + \theta_{\nu\mu} - 
u_{\mu} u^{\alpha}\nabla_{\alpha} u_{\nu}\,,
\end{equation} 
where 
\begin{eqnarray}
 \omega_{\mu\nu} &=& \frac{1}{2}(v_{\mu\nu}-v_{\nu\mu})\,, \label{vort}\\
 \theta_{\mu\nu} &=& \frac{1}{2}(v_{\mu\nu}+v_{\nu\mu})\,, \label{deform}
\end{eqnarray}
are the symmetric and anti-symmetric part of the projected four-velocity gradient (the vorticity and deformation tensors, respectively), and $v^{\mu\nu}=h^{\mu \alpha }h^{\nu \beta }u_{\alpha;\beta}$.
Note that if the electron four-velocity is hypersuperface orthogonal (and therefore the orthogonal planes are spacelike hypersuperfaces) the plasma is vorticity free and vice-versa.   
We also define the electric component of the EM tensor as $E^{\mu} \equiv 
F^{\mu}{}_{\nu}u^{\nu}$, the magnetic component as $B_{\mu\nu} \equiv h_{\mu}{}^{\alpha}h_{\nu}{}^{\beta}F_{\alpha \beta}$, and the Larmor 
tensor as ${\omega_{\rm L}}^{\mu\nu} = -\frac{e}{m_e} B^{\mu\nu}$. 
By differentiating and expanding to first order the system of equations~\eqref{eq:Maxwell}-\eqref{eq:normalisation}, we obtain a set of four differential equations for the perturbed quantities $\tilde{F}^{\mu\nu}$, $\tilde{u^\mu}$ and $\tilde{n}$. Using the momentum and Maxwell's equations, one can obtain an equation for the perturbed EM tensor $\tilde{F}^{\mu\nu}$, accounting for the influence of the 
gravitational potential and the moving plasma~\cite{1981A&A....96..293B},
\begin{align}\label{eq:master}
  h^\alpha{}_\beta u^\delta \nabla_\delta\nabla_\gamma\tilde{F}^{\beta\gamma}-\omega_{\rm pl}^2\tilde{F}^{\alpha\beta}u_\beta& \nonumber \\
 +(\omega^\alpha{}_\beta+{\omega_L}^\alpha{}_\beta+\theta^\alpha{}_\beta+\theta h^\alpha{}_\beta+\frac{e}{m_e}E^\alpha u_\beta)\nabla_\gamma\tilde{F}^{\beta\gamma}&=0,\nonumber\\
\end{align}
where $\theta = \theta^{\mu}{}_{\mu}$. 
The above equation is valid for any stationary background geometry, in particular also for a spinning (Kerr) BH spacetime surrounded by a cold and collisionless plasma. 
In the following, we analyse this equation in the Landau gauge, $\tilde{A}^{\mu}u_{\mu}=0$. 

As discussed in~\citetalias{Cannizzaro:2020uap}, in Minkowski spacetime Eq.~\eqref{eq:master} implies a dispersion relation with two solutions: the longitudinal modes, with frequency $\omega = \omega_{\rm pl}$, and the two transverse modes, with $\omega^2 = |k|^2 + \omega_{\rm pl}^2$~\cite{Raffelt:1996wa}. These dispersion relations are modified by the presence of collisions and finite-temperature effects -- even in the flat-spacetime limit -- in a nontrivial way, as discussed below.

Finally, note also that Eq.~\eqref{eq:master} contains third-order derivatives of the EM potential and, as discussed in~\citetalias{Cannizzaro:2020uap}, differs significantly from the Proca equation often used in the literature to model the effective mass of photons propagating in a
plasma.

\subsection{Collisional plasma}

In~\citetalias{Cannizzaro:2020uap} we have ignored particle collisions in the plasma. This amounts to assume that the collision rate between electrons and ions 
is much smaller than the characteristic oscillation frequency of the plasma, $\omega_{\text{pl}}$. In other words, it was assumed that the time $\tau$ between two electron-ion collisions is much longer than the other physical timescales in the problem.
Here we show that indeed the inclusion of collision effects does not affect our results in the astrophysical environments of interest. 

Let us first modify the set of Eqs.~\eqref{eq:Maxwell}-\eqref{eq:normalisation} in order to take into account electron-ion collisions. The electron equation of motion acquires an extra term~\cite{CovariantCollision}
\begin{equation}
    u^{\mu} \nabla_{\mu} u^{\nu} = \frac{e}{m_e} F^{\nu}{}_{\mu}u^{\mu} - \frac{1}{\tau}u^{\nu} .\label{eq:momentumCollision}
\end{equation}
At the microscopic level, $\tau$ can be thought as arising from Coulomb collisions between electrons and ions around the BH~\cite{Dubovsky:2015cca} \begin{equation}
    \tau \simeq \frac{2\pi \, m_e^2 v_e^3}{n_e e^4 \log \Lambda}\,,
\end{equation}
where $v_e$ is the typical electron velocity and $\log \Lambda$ is the Coulomb logarithm. 
However, in the interest of  generality, in the following we will treat the collision timescale $\tau$ as an independent parameter in the perturbed equations.

With this addition the perturbation equation~\eqref{eq:master} becomes
\begin{align}\label{eq:masterCollisions}
  h^\alpha{}_\beta u^\delta \nabla_\delta\nabla_\gamma\tilde{F}^{\beta\gamma}-\omega_{\rm pl}^2\tilde{F}^{\alpha\beta}u_\beta \boldsymbol{+ \frac{1}{\tau} h^\alpha{}_\beta \nabla_\gamma \tilde{F}^{\beta \gamma}}  \nonumber& \\
 +(\omega^\alpha{}_\beta+{\omega_L}^\alpha{}_\beta+\theta^\alpha{}_\beta+\theta h^\alpha{}_\beta+\frac{e}{m_e}E^\alpha u_\beta)\nabla_\gamma\tilde{F}^{\beta\gamma}&=0, \nonumber\\
\end{align}
where we highlighted the new term due to collisions in bold.

In Sec.~\ref{sec:numerical_c_t} we will solve this equation numerically for a nonrotating BH, and show that the effect of collisions can be safely neglected. As a back-of-the-envelope estimate, one can compare the collision timescale due to electron-proton Coulomb interactions with the other relevant timescale in the problem: $\omega_{\rm pl}^{-1}$. For densities and temperatures of astrophysical relevance, the collision timescale is much longer than the plasma oscillation time,
\begin{align}
\tau \, \omega_{\rm pl} 
\simeq 2 \times 10^{11}\, \left(\frac{T}{10^7 \rm K}\right)^{3/2} \, \left(\frac{n_e}{ 10^4 \rm cm^{-3}}\right)^{-1/2},
\label{eq:estimate_collisions}
\end{align}
where we estimated the typical electron velocity as their thermal velocity $v_e \simeq 0.03\sqrt{T/(10^7 \rm K})$ and took $\Lambda\simeq 20$. 
Notice that $\tau \omega_{\rm pl}\gg1$ even for much higher values of $n_e$, as those typical of accretion disks in the vicinity of the BH.
Furthermore, using the background electron velocity to estimate the collision timescale is a conservative choice. In fact, when electrons are accelerated to relativistic velocities (as in the case of the superradiant instability discussed below), the collision timescale is even longer. 

\subsection{Warm plasma}

So far, we have considered a cold plasma and ignored pressure terms in the electron equations of motion. We now turn our attention to thermal corrections arising in a warm plasma with temperature $T$. Thermal corrections are conceptually different from the relativistic and nonlinear corrections studied in Refs.~\cite{Cardoso:2020nst, Blas:2020kaa}, arising from the acceleration of electrons to very large velocities by a strong electric field. 
These accelerations, however, do not imply that the plasma temperature is high. In a warm (hot) plasma the \emph{thermal} velocity is comparable to (much higher than) the typical velocities of the propagating modes, which might not be the case even if the electrons are relativistic. Here we study what happens when thermal motion is turned on and the resulting thermal pressure gradients need to be included. 

The warm plasma model adopted here is an intermediate framework between the cold plasma model (where thermal motion is completely neglected) and the hot plasma model (where thermal motion is relevant and cannot be treated within the fluid description adopted here).
Since the velocities associated with the typical temperature of an accretion disk [Eq.~\eqref{Tdisk}] are smaller than, or at most comparable to, the phase velocity of the propagating EM mode described by a quasibound state (see~\citetalias{Cannizzaro:2020uap} and the estimate below), an intermediate, warm-plasma approximation is well justified.

In a warm-plasma model, the momentum equation of the electrons [Eq.~\eqref{eq:momentum}] must be modified with a pressure correction~\cite{Ahmedov:2010mt},
\begin{equation}
\label{eq:ThermalCorrections}
u^{\mu} \nabla_{\mu} u^{\nu} = \frac{e}{m_e} F^{\nu}{}_{\mu}u^{\mu}-\nabla^{\nu}p
\end{equation}
where $p=n k_B T$ is the pressure of an ideal gas. The system of equations must be closed by an equation of state $p=p(\rho)$ with $\rho = m_e n_e$ in the nonrelativistic regime.
In this case the equation for the perturbed EM field [Eq.~\eqref{eq:master}] is modified to
\begin{align}
\label{eq:EqThermal0}
  h^\alpha{}_\beta u^\delta \nabla_\delta\nabla_\sigma\tilde{F}^{\beta\sigma}-\omega_{\rm pl}^2\tilde{F}^{\alpha\beta}u_\beta +e\,\gamma\, v_{ \rm th}^2 \, h^\alpha{}_\beta \nabla^\beta \tilde{n}_e & \nonumber\\
 +(\omega^\alpha{}_\beta+{\omega_L}^\alpha{}_\beta+\theta^\alpha{}_\beta+\theta h^\alpha{}_\beta+\frac{e}{m_e}E^\alpha u_\beta)\nabla_\sigma\tilde{F}^{\beta\sigma}&=0\nonumber\,,\\
\end{align}
where $v_{\rm th}^2=k_B T/m_e$ is the electron thermal velocity and we assumed a polytropic equation of state with index $\gamma$, i.e., $p  \propto \rho^{\gamma}$. 
This allows us to relate, at leading order, a perturbation in the temperature $\tilde{T}$ to a perturbation in the electron density: $\tilde{T} = (\gamma - 1) T \, \tilde {n}_e/n_e$. 

Maxwell's equations also relate the density perturbation to the perturbation of the EM tensor. 
The thermal correction in Eq.~\eqref{eq:EqThermal0} can thus be expressed in terms of the EM tensor perturbation alone, 
\begin{align}
\label{eq:EqThermal}
 & h^\alpha{}_\beta u^\delta \nabla_\delta\nabla_\sigma\tilde{F}^{\beta\sigma}-\omega_{\rm pl}^2\tilde{F}^{\alpha\beta}u_\beta 
  \boldsymbol{-\,\gamma\, v_{\rm th}^2 \, h^\alpha{}_\beta \nabla^\beta u_\mu \nabla_\nu \tilde{F}^{\mu\nu}}  \nonumber \\
 &+(\omega^\alpha{}_\beta+{\omega_L}^\alpha{}_\beta+\theta^\alpha{}_\beta+\theta h^\alpha{}_\beta+\frac{e}{m_e}E^\alpha u_\beta)\nabla_\sigma\tilde{F}^{\beta\sigma}=0\,,\nonumber\\
\end{align}
where again the term due to thermal pressure modifying Eq.~\eqref{eq:master} is highlighted in bold. In the next section we shall solve this equation numerically in a Schwarzschild background and show that the effect of the plasma temperature can be safely neglected in realistic astrophysical settings, especially for the dominant quasibound states of the system. As a simple numerical estimate, Eq.~\eqref{eq:EqThermal} suggests that thermal effects will be negligible as long as 
\begin{equation}
\gamma v_{\rm th}^2 \simeq \gamma 10^{-2}\sqrt{\frac{T}{10^7 \rm K}}\ll 1\,.
\end{equation}

\section{Collisional and thermal corrections: numerical results}\label{sec:numerical_c_t}

\subsection{Harmonic decomposition and numerical method}\label{sec:numerical}

We begin by studying how the EM quasibound states computed in~\citetalias{Cannizzaro:2020uap} around a nonspinning BH are modified by collisions and thermal effects. In a Schwarzschild spacetime, the line element can be written as
\begin{equation}
    ds^2 = - f(r) dt^2 + f(r)^{-1} dr^2 + r^2 d\Omega_2^2,
\end{equation}
with $f(r) = 1- 2 M /r$, where $M$ is the BH mass.
In this background, both the vorticity and the deformation tensors are zero. We also assume that the plasma is unmagnetised, $B_{\mu\nu}=0$ (and therefore ${\omega_{\rm L}}^{\mu\nu}= 0$), and static. The latter assumption is justified by the fact that the accretion timescale is typically much longer than the dynamical timescale of the EM modes, see~\citetalias{Cannizzaro:2020uap}. The electron four-velocity in a static plasma 
is $u^\alpha=(u^0, \vec{0})$, with $ u^0=f(r)^{-1/2}$. The electric field has then only one nonvanishing radial component, $E^\alpha=(0, m_e/e \, (u^0)^2\Gamma^r_{00},0,0)$, where $\Gamma^\mu_{\alpha\beta}$ are the standard Christoffel's symbols of the Schwarzschild geometry. 

In any spherically-symmetric spacetime, it is convenient to separate the angular and radial parts of a field by performing a multipolar expansion. Following Ref.~\cite{Rosa:2011my}, we perfom the same decomposition as in~\citetalias{Cannizzaro:2020uap}. Namely, we introduce a basis of four-vector spherical harmonics,
\begin{align}
    Z_\mu^{(1)lm}&=c_1 \,[1,0,0,0]Y^{lm}(\theta, \phi)\label{eq:Sphericalh1} ,\\
    Z_\mu^{(2)lm}&=c_2 \,[0,f^{-1},0,0]\, Y^{lm}(\theta, \phi), \label{eq:Sphericalh2} \\
    Z_\mu^{(3)lm}&=c_3 r\,[0,0,\partial_{\theta},\partial_{\phi}] \, Y^{lm}(\theta, \phi), \label{eq:Sphericalh3}\\
    Z_\mu^{(4)lm}&=c_4 r\, [0,0,\sin^{-1}\theta\partial_\phi,-\sin\theta\partial_{\theta}] \, Y^{lm}(\theta, \phi) \label{eq:Sphericalh4}, 
\end{align}
where $c_1=c_2=1$, $c_3=c_4=1/\sqrt{l(l+1)}$ and $Y^{lm} $ are the standard scalar spherical harmonics. The vector spherical harmonics satisfy the orthogonality condition
\begin{equation}
    \int d\Omega Z^{(i)lm}_{\mu}\hat\eta^{\mu\nu}Z^{(i')l'm'}_{\nu}=\delta^{ii'}\delta^{ll'}\delta^{mm'} ,
\end{equation}
with 
$\hat\eta^{\mu\nu}={\rm diag}[1,f^2,1/r^2,1/(r^2\sin^2\theta)]$. The vector potential perturbation is then decomposed as
\begin{equation}
\label{eq:multexp}
    \tilde A_\mu(r,t,\theta,\phi)=\frac{1}{r}\sum_{i=1}^4\sum_{l,m}c_iu_{(i)}^{lm}(t,r)Z_\mu^{(i)lm}(\theta,\phi).
\end{equation}
Using this decomposition and working in the frequency-domain, $u_{(i)}^{lm}(t,r)=u_{(i)}^{lm}(r)e^{-i\omega t}$, we can separate the polar (even-parity) sector, described by the functions $u_{(1)}$, $u_{(2)}$ and $u_{(3)}$, from the axial (odd-parity) sector, described by the function $u_{(4)}$. This separation is a consequence of the spherical symmetry of the background.

An important advantage of the ansatz~\eqref{eq:multexp} is that the field equations can be put into a Schr\"odinger-like form with respect to the tortoise coordinate, $dr_*/dr = f(r)^{-1}$,
\begin{equation}
    \frac{d^2}{dr_*^{2}} u_{(i)}-V_{(i)}(r)u_{(i)}=0,
\end{equation}
where we have suppressed the $l$, $m$ indices and
where $V(r)$ is an effective potential.
In the following, we will solve these equations as an eigenvalue problem using a direct integration shooting method, as in~\citetalias{Cannizzaro:2020uap}. This method solves the differential equations by numerically integrating from the horizon to infinity, imposing suitable asymptotic conditions. 
The asymptotic behavior of the potential at the horizon allows for a superposition of ingoing and outgoing waves, but physical modes at the horizon must be purely ingoing. Therefore, the perturbation near the horizon is written as
\begin{equation}
\label{eq:boundaryh}
    u_{(i)}\sim e^{-k_+ r_*}\sum_n b_{(i)\, n} (r-2M)^n,
\end{equation}
where $k_{+}=\sqrt{V_{(i)}(r\to r_+)}$
and the coefficients $b_{(i)\, n}$ can be found  as a function of $b_{(i)\, 0}$ by solving the field equations perturbatively near the horizon. For a collisionless plasma around a Schwarzschild BH, $k_+ = i \omega$~\cite{Cannizzaro:2020uap}. The solution at infinity is also a superposition of ingoing and outgoing waves,
\begin{equation}\label{eq:boundaryinf}
u_{(i)}\sim B_{(i)}e^{-k_{\infty}r_* }+C_{(i)}e^{+k_{\infty}r_*},
\end{equation}
where $k_{\infty}=\sqrt{V_{(i)}(r\to \infty)}$. 
For a nonspinning BH surrounded by cold collisionless plasma, we found $k_{\infty}=\sqrt{\omega_{\rm pl}^2(r\to\infty)-\omega^2}$~\cite{Cannizzaro:2020uap}. 
The solution $B_{(i)}=0$ defines outgoing waves at infinity, i.e., the quasinormal modes of the problem, while the solution $C_{(i)}=0$ describes solutions that decay exponentially at infinity, i.e., the quasibound states. In this work we are interested in the latter, so we will impose $C_{(i)}=0$. 

The full Schr\"odinger-like equations for the EM perturbations $u_{(i)}$ are then solved by matching the two asymptotic expansions~\eqref{eq:boundaryh} and~\eqref{eq:boundaryinf} with $C_{(i)}=0$. This determines the eigenvalues of the problem, i.e., the complex quasibound frequencies $\omega=\omega_R+i\omega_I$. This method is extended in Sec.~\ref{sec:Kerr} to treat perturbations of a spinning BH.

\begin{figure}[ht]
\centering
\includegraphics[width=0.49\textwidth]{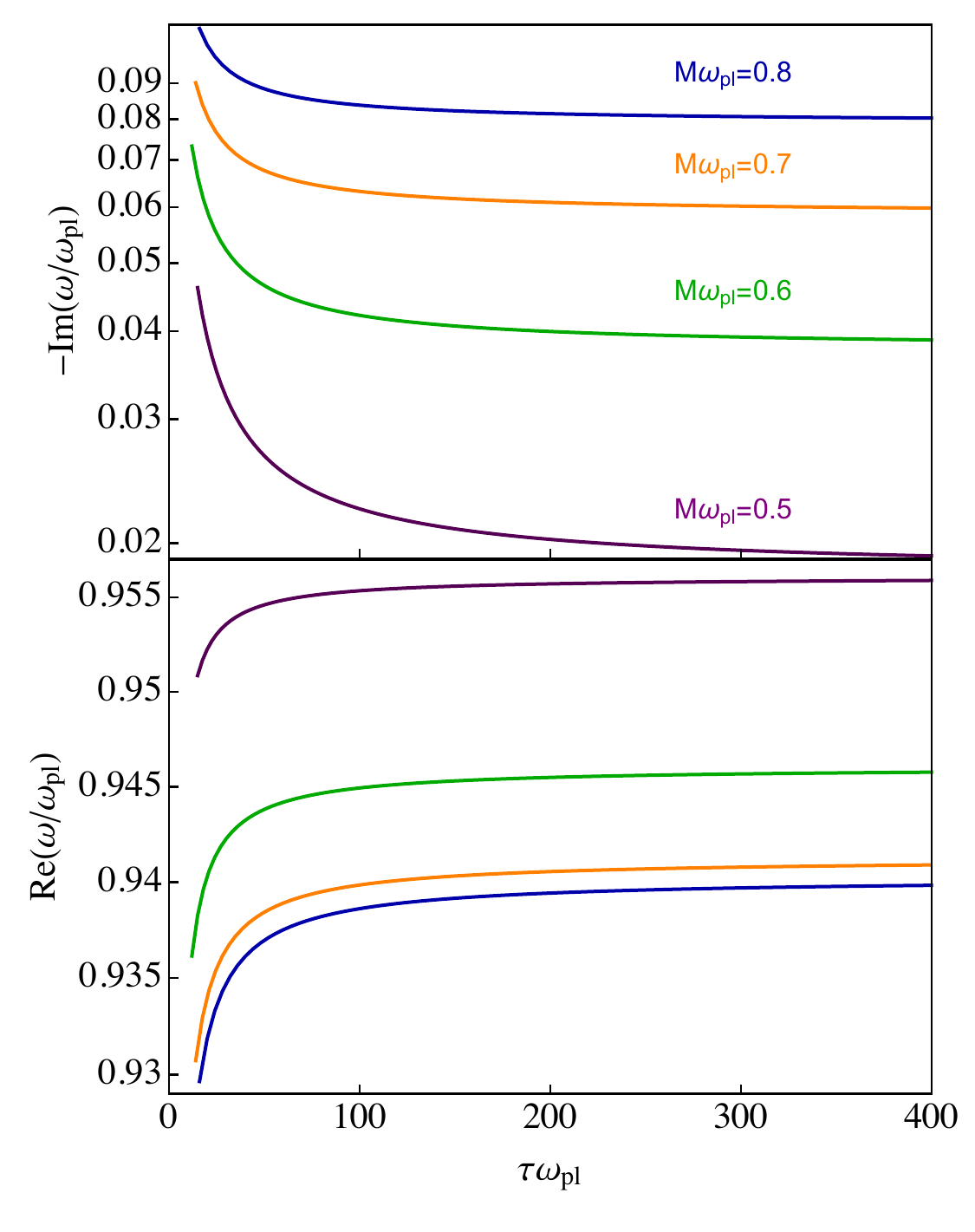}\hfill
\caption{Imaginary (top) and real (bottom) part of the axial $l=1$ mode for the quasibound states of a cold, collisional plasma in a Schwarzschild background as a function of the collision timescale $\tau$, normalized by the plasma frequency. For large collision time $\tau \gg 1/\omega_{\text{pl}}$, the results are identical to the collisionless case discussed in~\citetalias{Cannizzaro:2020uap}. When the collision time becomes very short, $\tau \ll 1/\omega_{\text{pl}}$, the collisions between electrons and protons shorten the lifetime of the bound states.  } 
\label{fig:axial_coll}
\end{figure}

\begin{figure}[ht]
\centering
\includegraphics[width=0.49\textwidth]{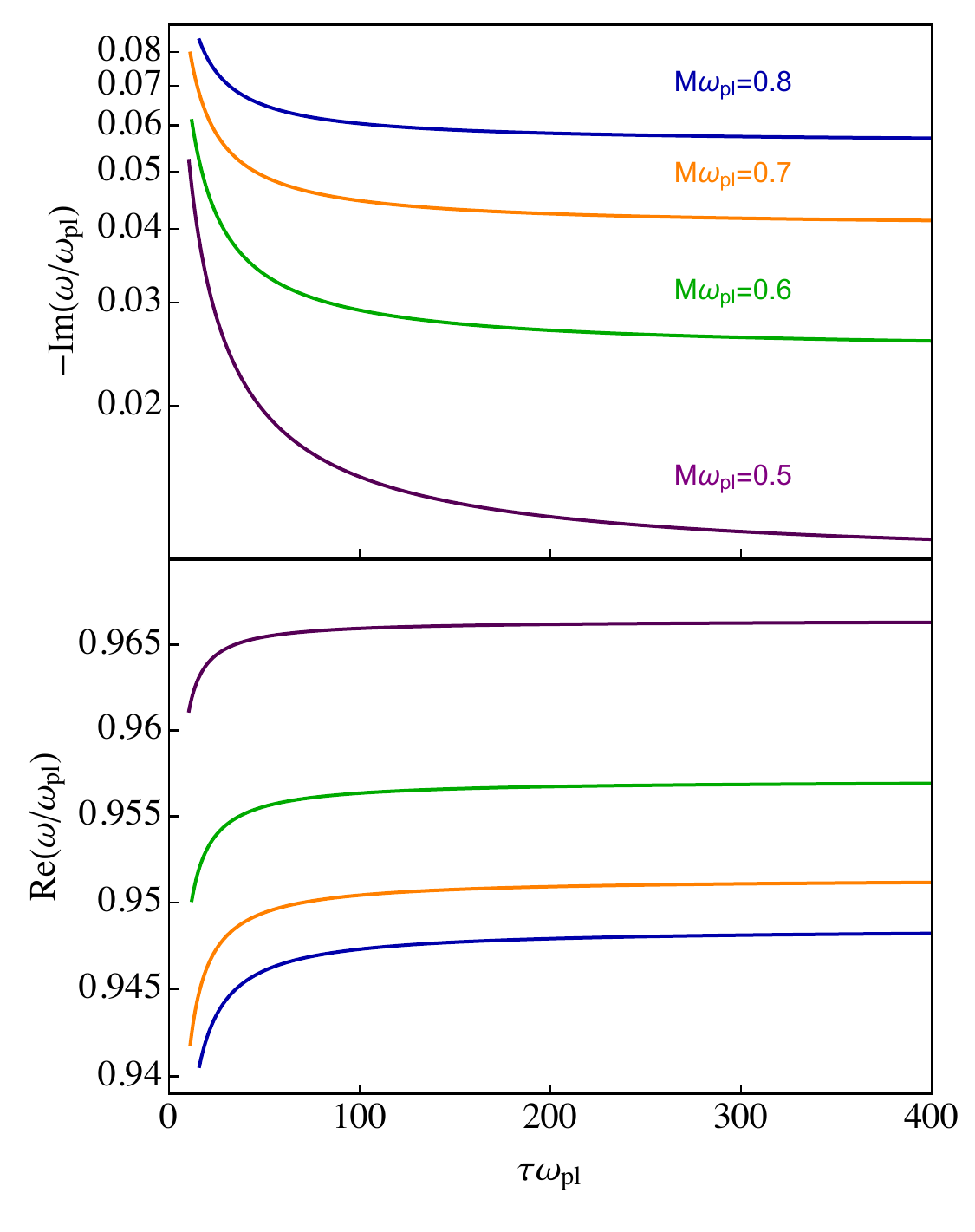}\hfill
\caption{ Same as in Fig.~\ref{fig:axial_coll} but for the polar sector. The behaviour of the modes is the same as in the axial case, and a quenching occurs as $\tau\omega_{\rm pl}\ll1$. } 
\label{fig:polar_coll}
\end{figure}

\subsection{Quasibound states in a collisional plasma}

We follow the procedure outlined above to solve for the EM perturbation in a collisional plasma, Eq.~\eqref{eq:masterCollisions}, around a nonspinning BH.
By performing the multipolar expansion~\eqref{eq:multexp}, Eq.~\eqref{eq:masterCollisions} gives
\begin{align}\label{eq:coll_u}
f^2 u_{(4)}''+2fMr^{-2} u_{(4)}' &\nonumber\\
+r^{-3}\left[r^3\omega^2-f r (l(l+1)+r^2\omega_{\rm pl}^2)\right]u_{(4)} &\nonumber\\ +\frac{f  \omega_{\rm pl}^2 }{1-i\tau\omega f^{-1/2}}u_{(4)}&=0,
\end{align}
for the axial sector, and
\begin{align}
f^2 u_{(3)}'' 
+ f F(r)   u_{(3)}' &\nonumber\\
+ \left( -\frac{f l (l+1)}{r^2} +\omega ^2+\frac{i \sqrt{f} \tau  \omega  \omega _{\text{pl}}^2}{1-i  \tau  \omega f^{-1/2} } \right) u_{(3)} &=0, 
\end{align}
for the polar sector, where $u_{(i)}'=\partial_r u_{(i)}$ and $F(r)$ is given in Appendix~\ref{app:polar_coll}. As expected, when $\tau\omega\to\infty$ we recover the equations governing the collisionless case studied in~\citetalias{Cannizzaro:2020uap}. Interestingly, the standard collisional term $i\tau\omega$ appearing in flat spacetime is modified by a redshift factor, $(1-2M/r)^{-1/2}$.

In Fig.~\ref{fig:axial_coll} and ~\ref{fig:polar_coll} we show the imaginary and real part of the fundamental axial and polar modes, respectively, as a function of the collision time, for four values of the plasma frequency. 
For any value of $\omega_p$ the dependence on $\tau$ is qualitatively the same. Namely, for very large collision timescales $\tau \gg \omega_{\text{pl}}^{-1}$, the frequencies coincide with the collisionless frequencies. As expected, collisions are irrelevant when the time between two collisions is much longer than the characteristic time of plasma oscillations. On the other hand, when the time between collisions is short, $\tau \ll \omega_{\text{pl}}^{-1}$, the absolute value of the imaginary part starts to increase, i.e., the lifetime of the mode is shortened. Collisions between electrons and protons in this case rapidly quench the quasibound states. However, as shown in Eq.~\eqref{eq:estimate_collisions}, the limit $\tau \ll 1/\omega_{\text{pl}}$ is never realized in astrophysical environments, and collisions can be safely neglected. 
Indeed, realistic values of $\tau\omega_p$ are much bigger than those shown in Fig.~\ref{fig:axial_coll}.

This numerical result can be understood analytically as follows. 
Eq.~\eqref{eq:coll_u} can be re-written to resemble the standard axial equation in the collisionless case, which coincides with the Proca axial equation, as
\begin{equation}
\label{eq:axial_const}
    \mathcal{D}^{\tau}_2 u_{(4)}(r)=0,
\end{equation}
where $\mathcal{D}_2^{\tau} \equiv   \frac{d^2}{dr_*^{2}}+ \omega^2- f \Big(\frac{l(l+1)}{r^2}+\mu_{\rm eff}^2\Big)$ has the same differential form as in~\citetalias{Cannizzaro:2020uap} but with the plasma frequency replaced by a collisions-dependent effective mass
\begin{equation}
    \mu_{\rm eff}^2=\omega_{\rm pl}^2\Big(1-\frac{1}{1-i\tau\omega f(r)^{-1/2}}\Big).
\end{equation}
The effective mass determines the behavior of the quasibound states at infinity, $k_{\infty}=\sqrt{\mu_{\rm eff}^2(r\to\infty)-\omega^2}$.
Clearly, in the limit $\tau \rightarrow \infty$, the effective mass tends to the plasma frequency, and the collisionless spectrum is recovered. In the opposite limit, $\tau \rightarrow 0$, the effective mass goes to zero -- unable to spatially confine the modes -- and consequently the quasibound spectrum is quenched. Similarly, in the polar case we found an effective mass at infinity
\begin{equation}
    \mu_{\rm eff}^2\sim \omega_{\rm pl}^2\Big(1-\frac{1}{1-i\tau\omega}\Big),
\end{equation}
which asymptotically coincides with the axial one. Thus, also in the polar sector the effective mass term at infinity goes to zero as plasma becomes strongly collisional, and the polar quasibound spectrum is quenched.

\subsection{Quasibound states in a plasma with thermal corrections}
We now turn our attention to thermal corrections. Applying the multipolar decomposition to Eq.~\eqref{eq:ThermalCorrections}, we obtain a system of differential equations for the mode functions in frequency domain,
\begin{align}
    \nonumber
    f^2 \gamma v_{\rm th}^2 r^4 u_{(2)}'' + \frac{3}{2} \gamma  (f-1) f r^3 v_{\text{th}}^2 u_{(2)}' &\\
    \nonumber -f^2 r^3 \left(\gamma   v_{\text{th}}^2-1\right) u_{(3)}' &\\
    \nonumber - r^2 \big[ f l(l+1)+\gamma/2  \left(6 f^2+f-3\right) v_{\text{th}}^2&\\
    \nonumber +r^2(f \omega _{\text{pl}}^2- \omega ^2)\big] u_{(2)}&\\
    +\frac{1}{2} \gamma  f (3 f+1) r^2 v_{\text{th}}^2u_{(3)}&=0 \, ,
    \label{eq:decomposedset2th}\\
    \nonumber f^2 r^3 u_{(3)}'' +(1 - f) f r^2 u_{(3)}'&\\
    \nonumber -f (-\gamma v_{\rm th}^2+1) l(l+1) r^2 u_{(2)}'&\\
    \nonumber + r \left [-f \gamma v_{\rm th}^2 l(l+1)  - f \omega_{\rm pl}^2 r^2 + r^2 \omega^2\right] u_{(3)} &\\
    -  l(l+1) r \left[\gamma v_{\rm th}^2-f (2 \gamma v_{\rm th}^2+1)\right] u_{(2)} &=0 \, ,
    \label{eq:decomposedset3th}\\
    \nonumber r^2 f^2  
        u_{(4)}''+r (1-f)  f u_{(4)}'&\\-\left[ f (l + l^2 + r^2 \omega_{\rm pl}^2) + r^2 \omega^2\right] u_{(4)} &=0\,. \label{eq:decomposedset4th}
\end{align}

In the polar sector we find the usual distinction between longitudinal $u_{(2)}$ and transverse $u_{(3)}$ modes. The former are non-dynamical in a cold plasma, but become propagating, energy-transporting modes (Langmuir waves in flat spacetime~\cite{2017mcp..book.....T}) due to the thermal corrections. 
Indeed, from Eq.~\eqref{eq:decomposedset2th} it is easy to see that the degree of freedom $u_{(2)}$ becomes dynamical only for $ v_{\rm th}\neq 0 $. To show that this mode behaves as a Langmuir wave, we take the flat spacetime limit: for $r \rightarrow \infty$ and in momentum space Eq.~\eqref{eq:decomposedset2th} reads
\begin{equation}
    \label{eq:Bohm}
    \omega^2 u_{(2)}=(\omega_{\rm pl}^2+k^2 \gamma v_{\rm th}^2)u_{(2)},
\end{equation}
which is the Bohm-Gross dispersion relation describing Langmuir modes in a warm plasma~\cite{2017mcp..book.....T}. In the same limit, the equation for the transverse mode~\eqref{eq:decomposedset3th} becomes
\begin{equation}
    \omega^2 u_{(3)}=(\omega_{\rm pl}^2+k^2)u_{(3)}.
\end{equation}
Therefore, this EM wave is unaffected by thermal corrections at infinity. This is a general feature of a warm plasma model in flat spacetime: transverse waves are unaffected by pressure~\cite{inan2010}. 
In curved spacetime, thermal corrections couple longitudinal and transverse polar modes and the dynamics is more involved.

On the other hand, and more importantly for our scopes, the axial sector governed by Eq.~\eqref{eq:decomposedset4th} is unaffected by first-order thermal corrections and is the same as in the cold-plasma case studied in~\citetalias{Cannizzaro:2020uap}. Indeed, being a transverse mode, $u_{(4)}$ is unaffected at infinity, and due to the spherical symmetry of the spacetime it does not couple to the polar, longitudinal, thermally-affected modes even near the BH. Given that the axial modes are the most relevant ones for the quasibound spectrum of the system, we conclude that warm-plasma corrections can be safely neglected when studying EM quasibound states and superradiance in astrophysical systems. 

Thermal effects should be carefully considered for a hot plasma, where transverse modes also become affected by temperature~\cite{inan2010}. In this case, the system must be studied using kinetic theory, solving a set of coupled Vlasov-Maxwell equations. A study of this type is beyond the scope of this work and left for future investigation.

\subsection{Polar quasibound states in the low plasma frequency regime: reflection point in inhomogeneous plasmas}\label{sec:resonance}

In ~\citetalias{Cannizzaro:2020uap}, we computed EM quasibound states in a cold, collisionless plasma at the linear level. For the polar sector, we were unable to explore the $\omega_{\rm pl}M\ll1$ regime with high precision, and we could only provide an estimate of the behaviour of the quasibound spectrum in this limit. In the following, we explain the physical origin of the issues encountered in the polar sector in the low-plasma-frequency limit, relating them to the behaviour of inhomogeneous plasmas in flat spacetime. Furthermore, we show that, as in flat spacetime, the inclusion of dissipative mechanisms such as collisions, thermal or nonlinear effects can ``smooth out'' the low-plasma-frequency regime.

The fact that, in curved spacetime, plasma behaves as an inhomogeneous medium -- even when assuming a constant electronic density -- can be understood by inspecting the monopole sector of the polar equation in the cold, collisionless case. As shown in~\citetalias{Cannizzaro:2020uap}, this reads
\begin{equation}
\label{eq:plasmons}
    (\omega^2-\omega_{\rm pl}^2 f ) u_{(2)}=0.
\end{equation}
This resembles the equation describing longitudinal modes $\psi_L$ in plasma physics: $\epsilon \psi_L=0$, where $\epsilon$ is the dielectric tensor\footnote{The dielectric tensor reduces to a scalar when the plasma is isotropic (unmagnetized), as assumed here.}. The solution to this equation gives the dispersion relation for longitudinal modes, $\epsilon=0$. By analogy, we introduce an effective dielectric tensor in the BH background, 
\begin{equation}
    \epsilon_{\rm eff}=1-\frac{\omega_{\rm pl}^2}{\omega^2}(1-\frac{2M}{r}).
    \label{eq:eff_dielectric}
\end{equation}
Note that in the near-horizon limit $r\sim 2M$ Eq.~\eqref{eq:plasmons} admits only the trivial solution $u_{(2)}=0$. Indeed, no electrostatic modes can exist at the BH horizon. Notice that the same occurs in the case of a massive (both scalar and vector) field, making the mass contribution subdominant at the horizon~\cite{PhysRevD.73.024009,Rosa:2011my,Price:1971fb}. 
In the flat spacetime limit, we recover the standard solution for electrostatic modes $\omega^2=\omega_{\rm pl}^2$, as discussed in~\citetalias{Cannizzaro:2020uap}. 

However, due to the BH curvature correction, the effective dielectric gains a dependence on the radius, $\epsilon_{\rm eff}=\epsilon_{\rm eff}(\omega,r)$. Therefore, plasma can be considered effectively \emph{inhomogeneous} even when $n_e={\rm const}$, due to the spacetime curvature~\cite{2015PhRvD..92j4031P}. A peculiar characteristic of inhomogeneous plasmas is the presence of a spatial point, known as \emph{reflection point}, where the dielectric tensor vanishes, i.e., $\epsilon(\omega, r)=0$. At this point, the longitudinal electric field and some components of the transverse magnetic and electric field diverge~\cite{nla.cat-vn322064, PhysRevLett.28.795}. Clearly, this divergence is not physical: dissipation mechanisms should be added to the theory to make the field equations well-behaved at the reflection point. 

Some of the most important dissipative channels are absorbtion via collisions (as discussed above) and the formation of plasma waves. When collisions are taken into account, the dielectric tensor acquires an imaginary part, the divergence is removed and a sharp (but finite) Breit-Wigner resonance appears instead\footnote{A system in which collisions play the role of the dominant dissipative channel is the Earth's ionosphere , where the dielectric tensor depends on the altitude $z$~\cite{nla.cat-vn322064}.}.
Thermal corrections can also cure the divergence near the point $\epsilon=0$ ~\cite{PhysRevA.11.679, HinkelLipsker1992AnalyticEF}.

Clearly, these two effects play a major role in a strongly collisional or warm plasma. If dissipative effects are not strong enough, although the divergence is removed, the maximum value of the electric field may still be very large. In this case, nonlinear effects will play the role of the dominant dissipative effect~\cite{Cardoso:2020nst}. A nonlinear treatment would take into account the motion of electrons due to the strong electric field, modifying the plasma density near the resonance point. For an analysis of this nonlinear effect in flat spacetime, see \cite{Dragila1981RelativisticLO, PhysRevLett.28.795}.

A similar situation arises in our system. The dielectric tensor of Eq.~\eqref{eq:eff_dielectric} makes its appearance in the denominator of the effective potential in the polar sector (see Appendix A in ~\citetalias{Cannizzaro:2020uap}).
In the case at hand, another important dissipation channel arises: the BH horizon. In the large plasma frequency regime ($M\omega_{\rm pl}>0.4$), dissipation through the horizon is sufficient to give rise to a complex mode frequency with a non negligible imaginary part $\omega_{\rm I}\lesssim\omega_{\rm pl}$, so that the dielectric tensor is complex and the field is well behaved. On the other hand, in the small plasma frequency regime, the BH horizon alone cannot quench the resonance and we have $\omega_{\rm I}\ll\omega_{\rm pl}$. Given that in the relevant astrophysical scenarios the plasma is cold and collisionless to very good approximation, we expect that nonlinearities will play the role of the dominant dissipative channel in this regime. Nonlinear effects are therefore crucial for the high-precision computation of the polar sector for $\omega_{\rm pl}M\ll1$. 

A nonlinear analysis of the system will appear in future work of this series. In the following, we characterize the role played by dissipative channels in the vicinity of the reflection point by focusing on a strongly collisional plasma as modelled in the previous sections. Although in the astrophysical systems of interest collisions can be neglected and the main dissipative effect is of nonlinear nature, we use this analysis as a jumping off point to clarify how the divergence is cured when the dielectric tensor becomes complex. 
\begin{figure*}[ht]
\centering
\includegraphics[width=0.79\textwidth]{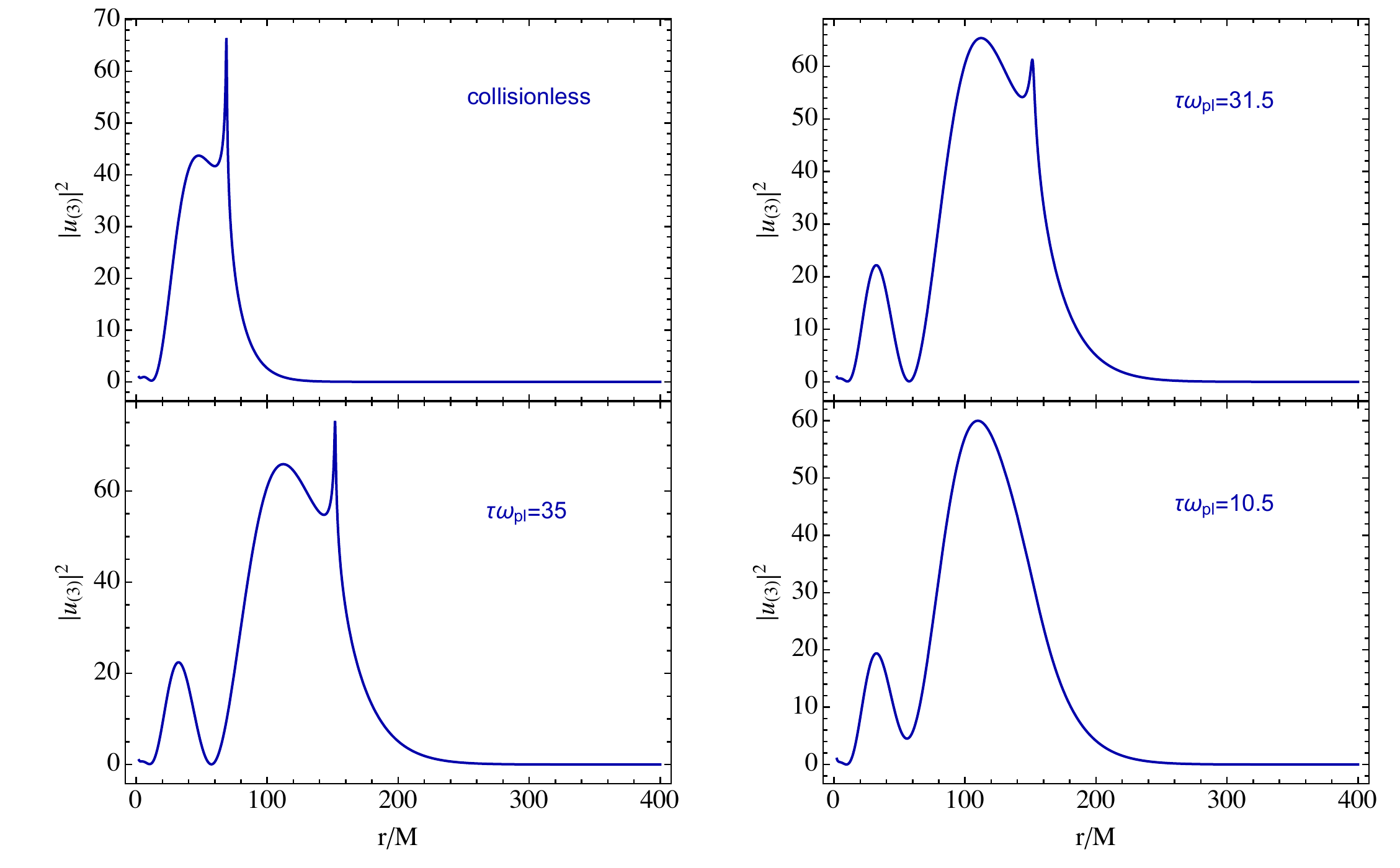}\hfill
\caption{Absolute value squared of the polar $l=1$ wavefunction $u_{(3)}$ for different values of the collision parameter $\tau$ at $M\omega_{\rm pl}=0.35$. As plasma becomes more collisional, the resonance at the reflection point is smoothed out. 
} 
\label{fig:reflect_coll}
\end{figure*}

Figure~\ref{fig:reflect_coll} shows the absolute value squared of the complex polar wavefunction $u_{(3)}$ for different values of the collision time $\tau$ for a small plasma frequency. The position of the sharp peak corresponds to a vanishing dielectric function, $r/M = 2/(1 - \omega^2/\omega_{\rm pl}^2)$ from Eq.~\eqref{eq:eff_dielectric}. 
The peak is ``smoothed out" as collisions come to dominate over the resonance, showing that, when dissipation channels are included in the theory, the field becomes well-behaved. In flat spacetime, the effect of nonlinearity can be similarly included by adding an effective collision frequency, $\nu_{\rm eff}$, to the dielectric tensor~\cite{Dragila1981RelativisticLO, PhysRevLett.28.795}. Thus, we expect that nonlinear effects should produce an effect similar to the one shown in Fig.~\ref{fig:reflect_coll}.

\section{Plasma-induced Superradiant instability: general formalism for linear perturbations}\label{sec:Kerr}

In previous sections we showed that collisions and thermal corrections have a small effect on quasibound states around a nonspinning BH. We are therefore justified to neglect these effects, and to focus our attention on plasma-induced superradiance in a Kerr background for a cold, collisionless plasma. In this section we derive the relevant linear perturbation equations to first order in the spin using a slow-rotation expansion~\cite{Pani:2013pma}. In the next section we will numerically solve these equations to find superradiant modes.

\subsection{Linearized plasma-photon dynamics in a Kerr spacetime}
\label{sub:spherical}

The line element of the Kerr metric in Boyer-Lindquist coordinates reads
\begin{align}\label{eq:Kerr}
    ds^2 =& - dt^2 + \Sigma (\frac{dr^2}{\Delta^2}+d\theta^2)+(r^2+a^2)\sin^2{\theta}d\phi^2 \nonumber \\& +\frac{2Mr}{\Sigma}(a \sin^2{\theta}d\phi-dt)^2 ,
\end{align}
where $\Delta(r)= r^2-2Mr+a^2$, $\Sigma(r,\theta)=r^2+a^2\cos^2{\theta}$ and $Ma$ is the angular momentum.

As in the nonspinning case, since the plasma accretion timescale is much longer than the dynamical timescale of the problem, we assume a static plasma.
Static observers exist outside the ergoregion of a spinning BH and their four-velocity reads $u^\alpha=(u^0, \vec{0})$, with $ u^0=g_{00}^{-1/2}$ so that Eq.~\eqref{eq:normalisation} is satisfied. At ${\cal O}(a/M)$ the ergosphere coincides with the outer horizon, $r_+=2M+{\cal O}(a^2/M^2)$, so that, to this order, static observers exist all the way to the BH horizon. 

Using the momentum equation~\eqref{eq:momentum} we obtain the background electric four-vector field giving rise to the static configuration, $E^\alpha=(0, m_e/e \, (u^0)^2\Gamma^r_{00},  m_e/e \, (u^0)^2\Gamma^{\theta}_{00},0)$, which possesses both a radial and an angular component. We also assume the plasma to be unmagnetized ${\omega_{\rm L}}_{\mu\nu}= 0$. 

Contrary to the Schwarzschild case, where both the vorticity and deformation tensors vanish, in the Kerr spacetime the rotation of the BH induces a nontrivial vorticity in the fluid. The vorticity tensor is antisymmetric and has two nonvanishing components, 
\begin{align}
 \omega_{r \phi } &= -\frac{a  \left(\Sigma -2 r^2\right) \left(a^2-\Delta +r^2\right)}{2 r \sqrt{\Sigma } \left(\Delta +\Sigma -r^2-a^2 \right)^{3/2}} \sin ^2\theta ,\\
 \omega_{\theta \phi } &= -\frac{a \Delta   \left(a^2-\Delta +r^2\right)}{\sqrt{\Sigma } \left(\Delta+\Sigma -r^2-a^2 \right)^{3/2}} \sin \theta  \cos \theta ,
\end{align}
which are nonzero already to first order in the BH spin.
The deformation tensor remains zero in Kerr.

It is again convenient to separate the angular sector of the field from the radial one through a multipolar expansion. While in Schwarzschild the axial and polar sectors can be fully decoupled from each other thanks to spherical symmetry, in Kerr the axial and polar perturbations with different index $l$ are coupled, making the field equations more challenging to solve~\cite{Pani:2013pma}.
Several methods were developed in order to solve the equations of a massive spin-1 field in a Kerr background, using a slow-rotation expansion~\cite{Pani:2012vp, Pani:2012bp}, analytical methods valid for ultralight Proca fields~\cite{Baryakhtar:2017ngi, Baumann:2019eav}, or numerically either without separability of the equations~\cite{Cardoso:2018tly} or using a recently-discovered separability technique~\cite{Frolov:2018ezx,Dolan:2018dqv}. 

In the following, owing to the complexity of the field equations for the problem at hand, we use the slowly-rotating approach to solve the EM perturbation equation~\eqref{eq:master} perturbatively. The field equations are expanded with respect to the dimensionless spin parameter $\tilde a=a/M\ll1$ around the nonspinning case $\tilde a=0$, and solved at different orders. This method was shown to perform well for Proca fields at second order, even for values of the spin close to extremality~\cite{Pani:2012bp}. In our case, we are mostly interested in whether the quasibound modes discussed above and in~\citetalias{Cannizzaro:2020uap} can turn unstable in the superradiant regime.

Using the multipolar expansion~\eqref{eq:multexp} and a frequency-domain representation, 
at first order in the spin the field equations assume the following form
\begin{align}
    u_{(1)}^{l}&=0\,, \label{eq:decomposedset1} \\
    A_{l}+Q_{l,m}[\tilde{A}_{l-1}+(l-1)B_{l-1}]&\nonumber\\+Q_{l+1,m}[\tilde{A}_{l+1}-(l+2)B_{l+1}]&=0, \label{eq:decomposedset2}\\
    l(l+1) \alpha_l-im\zeta_l +im\gamma_l &\nonumber\\-Q_{l,m}[(l+1)(\eta_{l-1}-(l-1)\delta_{l-1})]&\nonumber\\+lQ_{l+1,m}[\eta_{l+1}+(l+2)\delta_{l+1}]&=0, \label{eq:decomposedset3}\\
    l(l+1) \beta_l+im\eta_l+im\delta_l-Q_{l,m}[(l+1)(\zeta_{l-1}&\nonumber\\+(l-1)\gamma_{l-1})]+lQ_{l+1,m}[\zeta_{l+1}-(l+2)\gamma_{l+1}]&=0, \label{eq:decomposedset4}
\end{align}
where $Q_{l,m}=\sqrt{\frac{l^2-m^2}{4l^2-1}}$ while the quantities $A_i$, $B_i$, $\tilde{A}_i$, $\alpha_i$, $\beta_i$, $\zeta_i, \gamma_i, \delta_i$, and $\eta_i$ involve the mode functions $u_{(2)}^{i}$, $u_{(3)}^{i}$, and $u_{(4)}^{i}$ and their derivatives and are listed in Appendix~\ref{app:eqs}.
This set of equations has a similar schematic form as the one obtained in the Proca case at linear order in the BH spin~\cite{Pani:2012bp}. In particular, perturbations with a given parity and angular-momentum number $l$ only couple with perturbations of opposite parity and index $l\pm 1$ (as shown in Appendix~\ref{app:eqs}, the terms $A_i$, $\alpha_i$, $\zeta_i, \gamma_i$ are polar quantities, while $B_i$, $\tilde{A}_i$, $\eta_i$, $\beta_i, \delta_i$ are axial). 
Note that $A_l$, $\alpha_l$, and $\beta_l$ contain corrections proportional to $m \tilde{a}$, whereas all other functions are proportional to $\tilde a$.

As a generic property of the set of linear perturbations, the terms multiplied by $Q_{l,m}$ do not affect the spectrum at first order in the spin and can be neglected at this order~\cite{Kojima,Pani:2012bp,Pani:2012vp,Pani:2013pma}. This leads to an axial-led equation for $u^l_{(4)}$,
\begin{equation}
    l(l+1) \beta_l+im\eta_l+im\delta_l=0\,, \label{eq:decomposedset4bis}
\end{equation}
and a polar-led system of equations for $u^l_{(1)}$, $u^l_{(2)}$, and $u^l_{(3)}$,
\begin{align}
    u_{(1)}^{l}&=0\,, \label{eq:decomposedset1bis} \\
    A_{l}&=0, \label{eq:decomposedset2bis}\\
    l(l+1) \alpha_l-im\zeta_l+im\gamma_l&=0, \label{eq:decomposedset3bis}
\end{align}
The two sectors are decoupled and do not involve couplings between different-$l$ modes.

\subsubsection{Axial sector at first order in the BH spin}
Using the explicit form of the coefficients given in Appendix~\ref{app:eqs}, Eq.~\eqref{eq:decomposedset4bis} can be rewritten as
\begin{equation}
\label{eq:axial_constB}
    \mathcal{D}_2 u_{(4)}(r)-\frac{4amM\omega}{r^3}u_{(4)}=\frac{4mMa\omega_{\rm pl}^2(r-2M)}{l(l+1) r^4 \omega}u_{(4)},
\end{equation}
where we have suppressed the $l$ superscript and introduced the differential operator $\mathcal{D}_2 \equiv   \frac{d^2}{dr_*^{2}}+ \omega^2- f \left(\frac{l(l+1)}{r^2}+\omega_{\rm pl}^2\right)$. 
Equation~\eqref{eq:axial_constB} deviates from the axial equation of a Proca field at first order in the spin~\cite{Pani:2012bp} due to the presence of the term on the right-hand side. This correction is due to the vorticity tensor, which vanishes in the nonspinning case. Therefore, even at first order, we expect the spectrum to deviate quantitatively from that of a Proca field.

\subsubsection{Polar sector at first order in the BH spin}
The polar sector can be reduced, at first order in the spin, to a single second-order differential equation
\begin{equation}
\label{eq:boundarys}
    \frac{d^2}{dr_*^{2}} \psi-V(r)\psi=0,
\end{equation}
for an appropriately defined field variable $\psi$ (see Appendix~\ref{app:Polar sector}). As in the nonspinning case, although the original equations depended on two independent functions $u_{(2)}$ and $u_{(3)}$, the polar sector describes only one dynamical degree of freedom. The other degree of freedom does not propagate and remains electrostatic. 

In the Schwarzschild limit ($a\rightarrow 0$) the effective potential reduces to the one obtained in~\citetalias{Cannizzaro:2020uap}. In the spinning case, the potential depends on the azimuthal number $m$. Moreover, the boundary condition at the horizon is modified, since $V(r\rightarrow r_+)=-(\omega^2-2m\omega\Omega_H)$, where $\Omega_H $ is the angular velocity at the horizon of locally nonrotating observers at first order in $\tilde a$. Note that this factor coincides with the expected superradiant factor, $(\omega-m\Omega_H)^2$  at first order in the BH spin.

As discussed in Sec.~\ref{sec:resonance}, owing to the importance of nonlinear effects in the $\omega_{\rm pl} M\ll1$ regime, a nonlinear analysis of the polar sector in a Kerr spacetime is necessary to investigate its spectrum. Therefore, in the next section we shall focus only on the superradiantly unstable axial sector.

\section{Plasma-induced Superradiant instability: numerical results}
\label{sec:results}

In this section we numerically solve the differential equations derived in Sec.~\ref{sec:Kerr} at the linear order in the BH spin parameter. In order to do so, we need to specify the plasma density profile which determines the radial dependence of the effective photon mass. As in~\citetalias{Cannizzaro:2020uap}, we consider two different plasma profiles: a homogeneous density profile and a Bondi-like spherical accretion flow.

When the density is homogeneous, so is the plasma frequency, $\omega_{\rm pl}={\rm const}$. This approximation is not realistic, especially close to the BH, but it allows us to elucidate the structure of the equations.

We then consider a Bondi-like accretion model, which provides the radial dependence of the electron density (and therefore of the plasma frequency). This model is used to describe accretion onto spherically symmetric compact objects. The plasma frequency is written as
\begin{equation}
    \omega_{\rm pl}^2(r) = \omega_{\rm B}^2\Big(\frac{2M}{r}\Big)^{\lambda} + \omega_{\rm \infty}^2, \label{Bondi}
\end{equation}
where $\sqrt{\omega_{\rm B}^2+\omega_\infty^2}$ is the plasma frequency at the horizon (since in realistic settings $\omega_\infty\ll\omega_{\rm B}$, with a little abuse of notation we shall refer to $\omega_{\rm B}$ as the horizon plasma frequency). The slope $\lambda$ depends on the adiabatic index of the gas (e.g., $\lambda = 3/2$ for monoatomic species). The constant term $\omega_{\infty}$ is the asymptotic plasma frequency at infinity, i.e., the interstellar medium plasma frequency far away from the central BH.

We solve Eqs.~\eqref{eq:axial_const} and~\eqref{eq:boundarys} numerically with the shooting method described in Sec.~\ref{sec:numerical} and in~\citetalias{Cannizzaro:2020uap}. The boundary conditions (and in particular $k_{+}$ and $k_{\infty}$ in Eqs.~\eqref{eq:boundaryh}-\eqref{eq:boundaryinf}) are modified according to the behavior of the effective potential at the horizon and at infinity.

\subsection{Intermezzo: the hydrogenic spectrum for superradiantly unstable Proca modes in the Kerr metric}

It is useful to remind the reader of the known properties of the superradiant modes of a Proca field with a constant Stueckelberg mass $\mu=\hbar\omega_{\rm pl}$ around a Kerr BH. 

Massive vector particles in Kerr spacetime can populate gravitationally quasibound states described by a set of complex eigenfrequencies $\omega=\omega_{ R}+i\omega_{ I}$, where typically $|\omega_I|\ll|\omega_R|$. The binding energy of the mode is described by $\omega_R$, while $\omega_I$ is related to the decay rate of the mode due to dissipation of energy at the horizon only, since modes with $\omega_R<\omega_{\rm pl}(r\to\infty)$ cannot dissipate at infinity and are long-lived. If the superradiant condition $\omega_R<m\Omega_H$ is met, the decay rate will turn into a growth rate, leading to an exponential amplification of the mode due to extraction of rotational energy from the BH. In particular, since the modes are confined in the vicinity of the BH by the Proca mass, this amplification becomes a continuous process leading to a superradiant instability~\cite{Brito:2015oca}. The spectrum of a Proca field in Kerr spacetime in the Newtonian limit (i.e., as long as the Compton wavelength of the mode is much larger than the size of the BH) is hydrogen-like~\cite{Pani:2012vp,Pani:2012bp,Baryakhtar:2017ngi},
\begin{eqnarray}
 \omega_{ R} &\sim&\omega_{\rm pl}\left(1-\frac{(M\omega_{\rm pl})^2}{2(l+S+1+n)^2}\right)\,,  \label{hydrogenic}\\
 M\omega_{ I} &\sim& \gamma_{lS} (M\omega_{\rm pl})^{4l+2S+5}(\tilde{a}m-2r_+\omega_R) \,,\label{wIslope}
\end{eqnarray}
where $l$ is the total angular momentum of the state with spin projections $S=-1$, 0, 1 (with $S=0$ for axial modes and $S=\pm1$ for the two polarizations of polar modes), $n$ is the overtone number ($n=0$ for the longest-lived, fundamental mode), and $\gamma_{lS}$ are constants (given, e.g., in Ref.~\cite{Brito:2015oca}). The most unstable mode is the polar dipole with $S=-1$, $l=1$.
In the spinning case the imaginary part acquires a factor $\omega_I\propto(\omega_R-m\Omega_H)$, which depends on the BH angular velocity $\Omega_H$. Therefore, in the superradiant regime, $\omega_R<m\Omega_H$, the modes with the smallest slope in the static case (namely, the polar dipole with $S=-1$ for the Proca case) become the ones with the shortest instability timescale, $\tau_{\rm inst}=1/\omega_I$~\cite{Brito:2015oca}. This analytical approximation is in excellent agreement with exact numerical 
results in the Newtonian limit~\cite{Detweiler:1980uk,Dolan:2007mj,Baryakhtar:2017ngi,Cardoso:2018tly,Dolan:2018dqv,Baumann:2019eav,Brito:2013wya,Brito:2020lup}.

In the next section we compute the EM superradiant modes for the plasma-photon system and compare them with the above hydrogenic behavior of a Proca field.

\subsection{Constant density plasma}
%
\begin{figure}[t]
\centering
\includegraphics[width=0.49\textwidth]{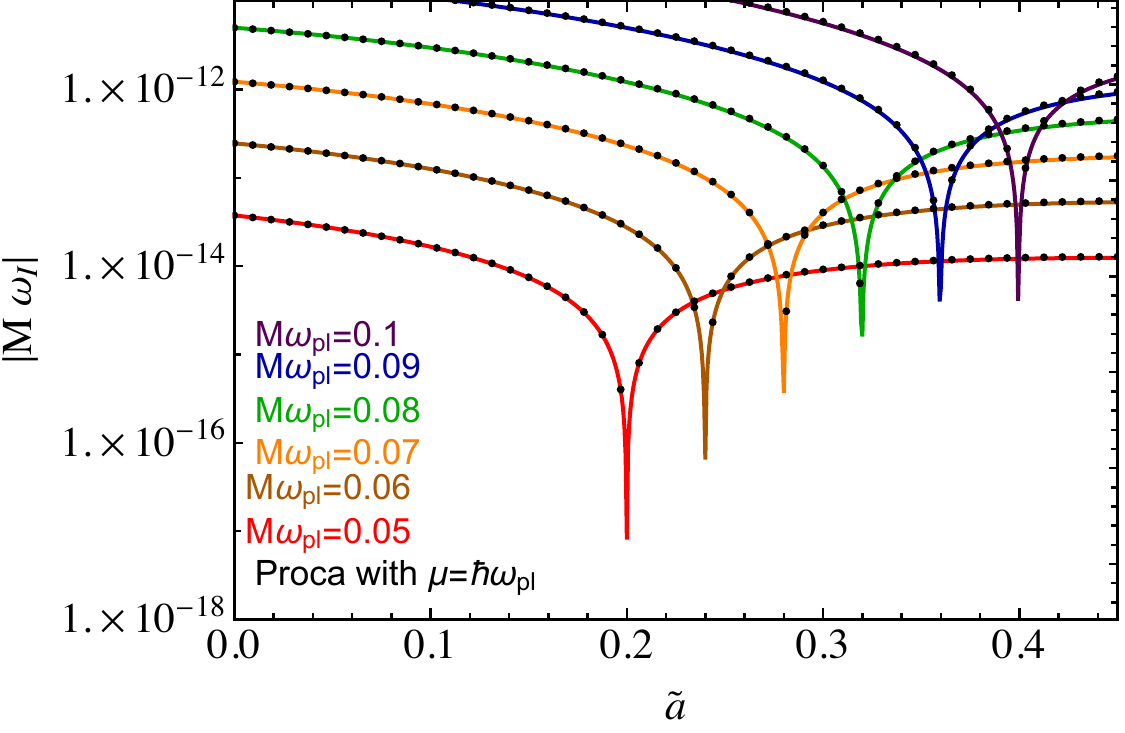}
\caption{Imaginary part of the axial $l=1= m$ mode for different values of the plasma frequency $\omega_{\rm pl}$. The imaginary part of the mode changes sign and becomes superradiant when the superradiant condition $\omega_{ R}<m\Omega_H$ is met. Black dots represent Proca modes with $l=1=m$. The difference between the two spectra is always less than $2\%$. }
\label{fig:Axial_Superradiant}
\end{figure}

In~\citetalias{Cannizzaro:2020uap}, we showed that the axial spectrum of the EM field around a nonspinning BH coincides with the axial sector of a Proca field, i.e. the real and imaginary part in the Newtonian regime coincide with Eqs.~\eqref{hydrogenic} and~\eqref{wIslope} with $S=0$.
On the other hand, the polar spectrum differed from Proca's and was subdominant in the large mass coupling regime $M\omega_{\rm pl}>0.4$.

Since in the slow-rotation expansion the spin is introduced perturbatively, we expect that the axial mode should provide the shortest instability timescale in the superradiant regime, at least for values of $M\omega_p$ which are not too small, in order to avoid the reflection point previously discussed for the polar modes.
 
In Kerr, the  field equations depend on the azimuthal number $m$. Therefore, the spectrum is characterised by a Zeeman-like splitting for different azimuthal numbers. In particular, modes with $m>0$ can become unstable if the superradiant condition $\omega_{\rm R}<m\Omega_H$ is satisfied.
Figure~\ref{fig:Axial_Superradiant} shows the absolute value of the imaginary part of the axial modes with $l=1=m$ at first order in the BH spin, for different values of the homogeneous plasma frequency. 
The black dots, for comparison, represent the modes of a Proca field with mass $\hbar \omega_p$. Interestingly, despite the fact that Eq.~\eqref{eq:axial_const} deviates from the axial Proca equation at first order, the two spectra almost coincide; the difference between an EM and a Proca mode at fixed $\omega_{\rm pl}$ is always less than $2\%$.  When the superradiant condition $\omega_{\rm R}<m\Omega_H$ is met, the imaginary part changes sign and the modes become superradiantly unstable. Since  $\omega_{\rm R}\sim \omega_{\rm pl}$ and $\Omega_H \sim a/(4M^2) +O(\tilde a^3)$ for a quasibound mode with $M\omega_{\rm pl}\ll 1$, the superradiant condition is met at $\tilde{a}\sim 4M\omega_{\rm pl}$, in good agreement with the crossing points in Fig.~\ref{fig:Axial_Superradiant}. 
Note that, since $\Omega_H={\cal O}(\tilde a)$, the superradiant condition requires $\omega_{\rm pl} M={\cal O}(\tilde a)$ or smaller, making it difficult to solve the equations numerically. Indeed, strictly speaking the superradiantly unstable modes require ${\cal O}(\tilde a^2)$ corrections (see Ref.~\cite{Pani:2012bp} for a discussion), although first-order results are already sufficiently accurate~\cite{Brito:2015oca}. 
When $M\omega_{\rm pl}\ll 1$, the real part of the modes depends only very weakly on the BH angular momentum and is very well approximated by the hydrogenic relation~\eqref{hydrogenic}, just as for Proca fields~\cite{Pani:2012bp}.

\subsection{Bondi Accretion Model}

We now analyze the spectrum for a Bondi accretion model, where the plasma frequency acquires a dependence on the radius $\omega_{\rm pl} \rightarrow \omega_{\rm pl}(r)$ as described by Eq.~\eqref{Bondi}. 
The modes can be obtained by solving Eq.~\eqref{eq:axial_const} taking into account the radial dependence of the effective mass.

\begin{figure}[th!]
\centering
\includegraphics[width=0.49\textwidth]{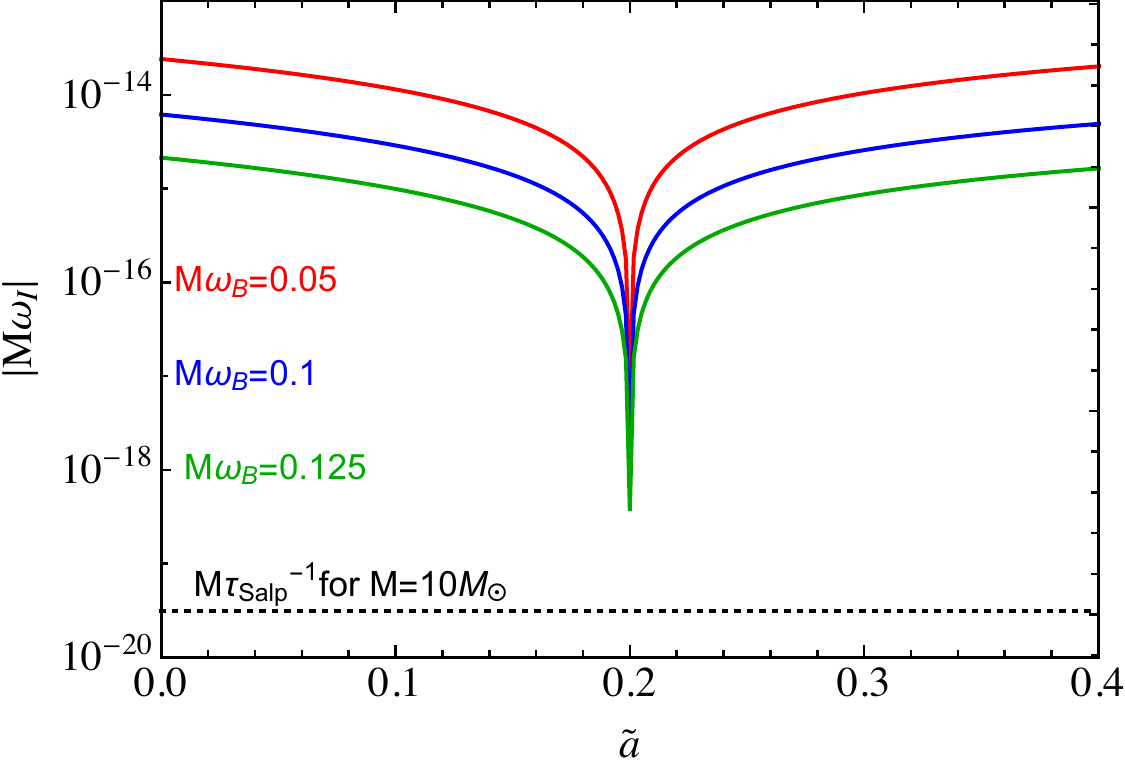}
\caption{Superradiant axial modes with $l=1=m$ in a Bondi accretion model [see Eq.~\eqref{Bondi}] with plasma asymptotic frequency $\omega_{\infty}=0.05/M$, for different values of the plasma frequency at the horizon, $\omega_{\rm pl}\approx\omega_{\rm B}$. The imaginary part of the modes is some orders of magnitude smaller than in the $\omega_{\rm pl}=const.$ case, and can become comparable to the Salpeter time (marked as a dashed line for the case of $M=10 M_\odot$).} 
\label{fig:Superradiant_Bondi}
\end{figure}

Figure~\ref{fig:Superradiant_Bondi} shows the (absolute value of the) imaginary part of the fundamental axial mode with $l= m=1$ as a function of the spin parameter $\tilde{a}$ for different values of $\omega_B$ in the case $\omega_{\infty}=0.05/M$. For these modes $\omega_R/\omega_{\infty} \sim 1$, therefore the superradiant condition $\omega_R<m\Omega_H$ becomes $4M\omega_{\infty}<\tilde{a}$, in agreement with the crossing points in Fig.~\ref{fig:Superradiant_Bondi}. Consequently, for sufficiently low plasma densities at the horizon, the system admits plasma-driven superradiant modes, with much larger timescales than in the $\omega_{\rm pl}=const.$ case. As shown in Fig.~\ref{fig:Superradiant_Bondi}, as $\omega_B$ increases the timescale of the mode also increases, and can become comparable to the Salpeter time. For such weakly-unstable modes, accretion must be taken into account; in particular, our formalism becomes inaccurate in this regime, as we assumed the plasma to be static. 

As shown in Ref.~\cite{Dima:2020rzg} (by using a Klein-Gordon toy model) and in~\citetalias{Cannizzaro:2020uap}, if the density at the horizon grows above a critical value, the spectrum is completely quenched, making the plasma-driven instability very fragile in realistic configurations.

\section{Conclusion}
\label{sec:discussion}

Studying the dynamics of an EM field propagating on a plasma is a complex and fascinating topic, already in flat spacetime. In this work, we have extended our ongoing analysis of the problem in curved spacetime, with the goal of understanding the confining role of plasma around an accreting BH.

In particular, we have extended our initial analysis~\cite{Cannizzaro:2020uap} of plasma-driven EM quasibound states in order to quantify the impact of thermal and collisional corrections in the plasma and the role of the BH spin. We showed that, for what concerns the existence of long-lived, quasibound EM states around an astrophysical BH, the plasma can be considered cold and collisionless with excellent approximation. 
We also showed that, when $M\omega_{\rm pl}\ll1$, the polar sector generically features a spatial reflection point, due to resonant properties akin to those of inhomogeneous plasmas in flat spacetime.
Dissipation channels such as collisions and nonlinearities can have a major impact on these resonances and are in fact needed to cure the singularities of the EM field at the reflection point.

We further demonstrated that the quasibound states can turn unstable when the BH rotates above the superradiance threshold. 
In particular, even in this case the axial sector can be described by a Klein-Gordon toy model with excellent approximation, while for the polar sector a nonlinear analisys is required. 

We limited our analysis to the case of linear perturbations of the EM field and plasma quantities. However, nonlinear effects in the plasma-photon interaction can quench the superradiant instability~\cite{Cardoso:2020nst}.
Nonlinearities are also the dominant dissipation mechanism curing the resonances in the polar sector in the astrophysical systems of interest. The next step will be to investigate the role of nonlinearities in detail by either numerically solving the full systems of equations, or by introducing effective nonlinear corrections.

We only considered nonrelativistic electrons, but relativistic corrections --~which might also quench the superradiant instability~\cite{Blas:2020kaa}~-- can be accommodated in our framework.
Other interesting extensions of this work include considering a magnetized plasma and studying the photon-plasma interactions around a spinning BH for generic spins.

\begin{acknowledgments}
%
P.P.~acknowledges financial support provided under the European Union's H2020 ERC, Starting 
Grant agreement no.~DarkGRA--757480. We also acknowledge support under the MIUR PRIN and FARE programmes (GW-NEXT, CUP:~B84I20000100001), and from the Amaldi Research Center funded by the MIUR program ``Dipartimento di Eccellenza'' (CUP:
B81I18001170001). A.C.~is supported by the Foreign Postdoctoral Fellowship Program of the Israel Academy of Sciences
and Humanities. A.C.~also acknowledges support from the Israel Science Foundation (Grant 1302/19), the US-Israeli BSF (Grant 2018236) and the German-Israeli GIF (Grant I-2524-303.7).
\end{acknowledgments}

\appendix

\section{Polar sector with collisions} \label{app:polar_coll}
The polar sector in the presence of collisions on a nonspinning background depends on the function $F(r)\equiv G(r)/H(r)$, where
\begin{widetext}
\begin{align}
G(r) = &f \tau  \omega _{\text{pl}}^2 \left[(3 f+1) f^{3/2} l (l+1)-2 i (f+1) f l (l+1) \tau  \omega +2 i (f-1) f r^2 \tau  \omega  \omega _{\text{pl}}^2-4 i (f-1) r^2 \tau  \omega ^3+4 (f-1) \sqrt{f} r^2 \omega ^2\right] \nonumber \\ &
+2 i \omega  (\tau  \omega +i \sqrt{f})^2 \left[2 f^2 l (l+1)+(f-1) r^2 \omega ^2\right],\\
H(r) = & 2 r \left[f \tau  \omega _{\text{pl}}^2-\omega  (\tau  \omega +i \sqrt{f})\right] \left[(\sqrt{f}-i \tau  \omega ) \left(f l (l+1)-r^2 \omega ^2\right)-i f r^2 \tau  \omega  \omega _{\text{pl}}^2\right] .
\end{align}
\end{widetext}

\section{Field decomposition in Kerr} \label{app:eqs}

In the following, we list the coefficients appearing in the decomposition of the plasma equations in a Kerr background, Eqs. ~\eqref{eq:decomposedset2},~\eqref{eq:decomposedset3} and ~\eqref{eq:decomposedset4},
\begin{align}
   \tilde{A}_l=&0, \label{eq:Atl}\\
    \eta_l=&-\frac{2 a l (l+1) M \omega ^2}{(r (r-2 M))^{3/2}} u_{(4)},\label{eq:etal}
\end{align}
\begin{widetext}
\begin{alignat}{2}\label{eq:Al}
  A_l={}& \frac{i }{r^{11/2} (r-2 M)^{5/2}} \big[&& -2 a m M r (M-r) (2 M-r)^2 u_{(3)}'' \nonumber\\&
  &&+ (2 M-r) \left(2 a m M \left(2 M (M-r)-r^4 \omega ^2\right)+l (l+1) r^4 \omega  (2 M-r)\right) u_{(3)}'\nonumber\\&
  &&+2 a m M (2 M - r) r (2 M^2 - 3 M r + r^2)u_{(2)}'
  +2 a m M r^3 \omega ^2 (r-M) u_{(3)}\nonumber\\&
  &&+ (r-2 M) \left(4 a m M \left(M^2-M r-r^4 \omega ^2\right)-r^4 \omega  (r-2 M) \left(l^2+l+r^2 \omega _{\text{pl}}^2\right)+r^7 \omega ^3\right) u_{(2)}
  \big],
\end{alignat}
\begin{alignat}{2}\label{eq:Bl}
    B_l={}& -\frac{2 a M}{r^{11/2} (r-2 M)^{5/2}} &&{}\big[
    r (2 M-r) \left(2 M^2-3 M r+r^2\right) u_{(4)}''
    + (2 M-r) \left(-2 M^2+2 M r+r^4 \omega ^2\right)  u_{(4)}'\nonumber\\& 
    &&+  (M-r) \left(l^2 (2 M-r)+l (2 M-r)+r^3 \omega ^2\right)  u_{(4)}
    \big],
\end{alignat}
\begin{alignat}{2}\label{eq:alphal}
   \alpha_l={}& \frac{i \omega }{r^{3/2} (r-2 M)^{3/2}} \big[&&
   -r (r-2 M)^2 u_{(3)}''
   + 2 M (2 M-r) u_{(3)}'
   + \left( r (r-2 M)^2 u_{(2)}' +2 M (r-2 M) \right)u_{(2)}\nonumber\\&
   &&+ \omega  \left(2 a m M-r^3 \omega \right)+r^2 (r-2 M) \omega _{\text{pl}}^2 u_{(3)}
   \big],
\end{alignat}
\begin{alignat}{2}\label{eq:zetal}
    \zeta_l={}&-\frac{2 a M}{r^{9/2} (r-2 M)^{3/2}} \big[&&
    -l (l+1) (M-r) (2 M-r) u_{(3)}'
    - r^4 \omega ^2 (2 M-r) u_{(2)}' 
    -l (l+1) r^3 \omega ^2 u_{(3)}\nonumber\\&
    &&+\left( l^2 \left(2 M^2-3 M r+r^2\right)+l \left(2 M^2-3 M r+r^2\right)+r^3 \omega ^2 (3 M-2 r)    \right) u_{(2)}
    \big],
\end{alignat}
\begin{align}\label{eq:deltal}
    \delta_l={}& -\frac{4 a M}{r^{9/2} (r-2 M)^{3/2}} \big[
    r (r-2 M)^2 u_{(4)}''
    + 2 M (r-2 M) u_{(4)}'
    + \left( l^2 (2 M-r)+l (2 M-r)+r^3 \omega ^2 \right)u_{(4)}
    \big],
\end{align}
\begin{align}\label{eq:gammal}
    \gamma_l={}& \frac{4 i a m M}{r^{9/2} (r-2 M)^{3/2}} \big[
    r (r-2 M)^2 u_{(3)}''
    + 2 M (r-2 M) u_{(3)}'
    -r (r-2 M)^2 u_{(2)}'
    + r^3 \omega ^2 u_{(3)}
    + 2 M (2 M-r) u_{(2)}
    \big],
\end{align}
\begin{alignat}{2}\label{eq:betal}
    \beta_l={}& -\frac{i \omega }{(r (r-2 M))^{3/2}} \big[&&
    -r (r-2 M)^2 u_{(4)}''
    +2 M (2 M-r) u_{(4)}'\nonumber\\&
    &&+ \left( \omega  \left(2 a m M-r^3 \omega \right)+l^2 (r-2 M)+l (r-2 M)+r^2 (r-2 M) \omega _{\text{pl}}^2 \right) u_{(4)}
    \big].
\end{alignat}
\end{widetext}
\section{Polar sector in Kerr}\label{app:Polar sector}

In the following, we outline the procedure to derive the polar potential from Eqs.~\eqref{eq:decomposedset2}, \eqref{eq:decomposedset3}, by neglecting the terms multiplied by $Q_{l,m}$ and using the expressions given in Appendix \ref{app:eqs}. From Eq.~\eqref{eq:decomposedset2}, it is possible to obtain an expression for $u_{(2)}'$ as a function of $u_{(2)}, u_{(3)}, u_{(3)}'$ and $u_{(3)}''$. This expression can be then inserted in Eq.~\eqref{eq:decomposedset3}, obtaining an equation that only contains $u_{(2)}$, $u_{(3)}$, $u_{(3)}'$ and $u_{(3)}''$. Thus, by solving this equation, it is possible to write $u_{(2)}$ as a function of the degree of freedom $u_{(3)}$ and its derivatives, 
$ u_{(2)}(r)={\cal F}[ u_{(3)}, u_{(3)}', u_{(3)}'']$.
Inserting this in Eq.~\eqref{eq:decomposedset3} allows us to obtain an equation for the decoupled variable $u_{(3)}$, which in general contains third-order radial derivatives. However, the latter are ${\cal O}(\tilde a^2)$ and can therefore be neglected to linear order in the BH spin. The resulting equation is a second-order equation for the decoupled variable $u_{(3)}$ in the form $a_2(r) u_{(3)}'' + a_1(r) u_{(3)}' + a_0(r) u_{(3)} =0$, where the coefficients $a_i$ are functions of $r$ and are at most linear in the BH spin. As in the nonspinning case, therefore, only one degree of freedom is propagating in the polar sector.

The differential equation for $u_{(3)}$ can also be written in a Schr\"odinger-like form through a variable redefinition and in terms of the tortoise coordinate.
At first order in the spin, $V\to-(\omega^2-2m\omega \Omega_H)$ near the horizon and $V\to \omega_{\rm pl}^2-\omega^2$ at infinity.

\bibliographystyle{utphys}
\bibliography{Ref}

\end{document}